\documentclass[twoside,11pt]{article}
\def\be{\begin{equation}}
\def\ee{\end{equation}}
\def\ba{\begin{align}}
\def\ea{\end{align}}
\def\bea{\begin{eqnarray}}
\def\eea{\end{eqnarray}}

\def\ba{\bar{\mathcal A}}

\usepackage{amsmath,amssymb,amsfonts}
\usepackage{amsfonts}
\usepackage{graphicx}
\usepackage{graphics}
\usepackage{breqn}
\begin{document}
\setlength{\unitlength}{10mm}

\title{Fundamental Symmetries of the Modified Anyonic Particle}

\author{
Salman Abarghouei Nejad
\\
{\scriptsize Department of Physics,
}\\{\scriptsize University of Kashan, Kashan 87317-51167, I. R. Iran}
\\
\\
Mehdi Dehghani
\\
{\scriptsize Department of Physics, Faculty of Science,
}\\{\scriptsize Shahrekord University, Shahrekord, P. O. Box 115, I. R. Iran}
\\
\\
Majid Monemzadeh
\\
{\scriptsize Department of Physics,
}\\{\scriptsize University of Kashan, Kashan 87317-51167, I. R. Iran}}

\date{}
\maketitle
\vspace{1cm}
\vspace{-1.5cm}\begin{abstract}
\noindent  
We try to increase the fundamental symmetries of the anyonic particle with the help of the symplectic formalism of constrained systems and gauging the model. The main idea of this approach is based on the embedding of the model in an extended phase space. After the gauging process had done, we obtain generators of gauge transformations of the model. Finally, by extracting the corresponding Poisson structure of all constraints, we compare the effect of gauging on the the phase spaces, the number of  physical degrees of freedom, and canonical structures of both primary and gauged models.



\end{abstract}
\newpage

\newcommand{\f}{\frac}

\newtheorem{theorem}{Theorem}[section]
\newcommand{\sta}{\stackrel}
\section*{Introduction}

The prototype Lagrangian for a free relativistic particle with mass $ m $, can be considered as similar to its nonrelativistic congener, i.e. $ L=\frac{1}{2}m \dot{x}^{2} $. Apparently, this Lagrangian does not apply any restriction on the particle to obey the second principle of relativity, i.e. having the velocity less than the speed of light. To remove such an ambiguity, we can use the Lagrangian of the free relativistic particle as,
\begin{equation}\label{L000}
L=\sqrt{p^{2}(\dot{x})+m^{2}}-m.
\end{equation}
Although this Lagrangian satisfies the second law of relativity, it depends only on the momenta (phase space coordinates) and must be transformed to the configuration space (i.e. to be a function of coordinates and corresponding  velocities), using a proper map. 
Replacing the relativistic momentum with the four velocity of the particle, we obtain the appropriate Lagrangian for the free relativistic particle \cite{Deriglazov}.

Also, the Lagrangian \eqref{L000} dose not include any spin degree of freedom. In classical mechanics point of view, this degree of freedom is not an observable quantity, but it will be emerged after quantization is applied. The spin can be classically added to the model with the help of a space-like vector $ n^{\mu} $ which satisfies the nonphysical condition, $ n^2+1=0 $.  A proposed scalar relativistic Lagrangian is,
\begin{equation}\label{L00}
L=\frac{m(\dot{x}. \dot{n})}{\sqrt{\dot{n}^{2}}}.
\end{equation}
This Lagrangian describes a particle with an arbitrary spin degree of freedom, which is called anyon \cite{chai1,anyon1,anyon11,anyon2,anyon3}. Due to its interesting statistics, it has been used to modelling different phenomena in physics, such as fractional quantum Hall effect, high-Tc superconductivity, and even describing some physical processes in the presence of cosmic strings  \cite{chai1}.

\section{Phase space of the Anyonic Model}
Now, we want to increase the gauge symmetries of this Lagrangian, in order to investigate whether the classical form of the spin degree of freedom will contain the physical property. Our tool to do such an investigation is the symplectic formalism \cite{FJ1, G2, FJ2, FJ3,FJ4}.

Using the definition of the canonical momentum, we will have,
\begin{align}\label{momenta0}
& p_{\mu}=\frac{m \dot{n}_{\mu}}{\sqrt{ \dot{n}^{2}}}, \nonumber \\
&\wp_{\mu}=m\frac{\dot{x}_{\mu}}{\sqrt{ \dot{n}^{2}}}-\frac{m\dot{n}^{2} \dot{x}_{\mu} }{( \dot{n}^{2})^{\frac{3}{2}}}.
\end{align}
which, $ p_{\mu} $ and $ \wp_{\mu} $ are conjugate momenta of $ x_{\mu} $ and $ n_{\mu} $ respectively.
The space-like condition of the added spin vector, imposes the following primary constraint as,
\begin{equation}\label{FC1}
\phi_{0}=n^{2} +1,
\end{equation}
which is recognised as a first class constraint, afterwards.

Also, the definition of momenta obtained in \eqref{momenta0}, produces some identities as follows, which depend on phase-space coordinates. These identities are called constraints and defined on the constraint surface.
\begin{eqnarray}
&& \phi_{1}=p^{2}-m^{2}, \label{PSC01}\\
&& \phi_{2}=n.p,  \label{PSC02}\\
&& \phi_{3}=\wp . p \label{PSC03}  .
\end{eqnarray}

Considering the Dirac's classification of constraints, we see that $ \phi_{0} $ and $ \phi_{1} $ are first class constraints, while $ \phi_{2} $ and $ \phi_{3} $ are second class ones \cite{3}. These second class constraints help us to describe the particle only with $ (x_{\mu},n_{\mu}) $ and their corresponding momenta with the Poisson structure which is obtained via Dirac's brackets as follows.
\begin{eqnarray}\label{Poisson0}
&& \{x_{\mu},x_{\nu}\}^{*} =\frac{1}{m^{2}}(n_{\mu} n_{\nu}-\wp_{\mu}\wp_{\nu}) \nonumber \\
&&  \{x_{\mu},p_{\nu}\}^{*} = \delta_{\mu\nu} \nonumber \\
&&  \{x_{\mu},n_{\nu}\}^{*} =\frac{1}{m^{2}}( n_{\mu}-p_{\mu}p_{\nu}) \nonumber \\
&&  \{x_{\mu},\wp_{\nu}\}^{*} =\frac{-1}{m^{2}}( n_{\mu}p_{\nu}+p_{\mu}) \nonumber \\
&&  \{n_{\mu},\wp_{\nu}\}^{*} =\delta_{\mu\nu}+\frac{p_{\mu}p_{\nu}}{m^{2}} \nonumber \\
&&  \{n_{\mu},n_{\nu}\}^{*} =\frac{-p_{\mu}p_{\nu}}{m^{2}} \nonumber \\
&&  \{\wp_{\mu},\wp_{\nu}\}^{*} =\frac{p_{\mu}p_{\nu}}{m^{2}} 
\end{eqnarray}
Obviously, this phase space is a noncommutative one in quantization \cite{Meljanac}.

While, the first class constraint $ \phi_{0} $ confirms the existence of a gauge symmetry in the model. As a matter of fact, we desire to increase those symmetries. Nevertheless, we want to check that this symmetry enhancement will be gained only with the help of $ n_{\mu} $, or there would be existed some new degrees of freedom for the particle. 

Indeed, having a second class constraint in the model indicates the presence of the redundant degree of freedom which spoils gauge symmetry \cite{park}. 

\section{Gauging the Anyonic Model}
In order to gauge the model, we must be assured that the canonical Hamiltonian is existed. But, we see that for this model,
\begin{align}
H_{c}=\dot{x}.p+\dot{n}.\wp -L,
\end{align}
which is vanished exactly.

Hence, the embedding procedure to enhance the gauge symmetry of the model is not applicable any more. The main reason is that this model is a mixed model, consisting of both first class and second class constraints.

To surmount the difficulty, we can use some auxiliary coordinates, such as auxiliary conjugate variables like $( \xi ,\pi ) $, to extend the phase space of the original model and convert a mixed physical system to a pure second class model, using the following extensions \cite{9,10},
\begin{align*}
& p_\mu \rightarrow p_\mu+\xi_\mu, \\
& H_c \rightarrow H_c+\frac{1}{2}\pi^{2},
\end{align*}
and transform the Lagrangian \eqref{L00}, using a gauge fixing term in the following replacement,
\begin{equation}\label{Lxi}
L \rightarrow L-\xi. \dot{x} +\frac{1}{2}  \dot{\xi}^{2}.
\end{equation}
Since this replacement is a gauge fixing term which is inserted in the gauge invariant Lagrangian, the arbitrariness of the gauge dependent variables will be destroyed and via the variation of the new Lagrangian, we obtain same equations of motion for the gauge invariant quantities.

It has been shown that the added variables and their corresponding momenta can be eliminated at the end of the gauging process, if they are second class in comparison to other constraints, using their constrained equation \cite{NC4}.

Thus, the Lagrangian \eqref{L00} will be modified to,
\begin{equation}\label{L(0)}
L^{(0)}=m^{2}\frac{\dot{x}. \dot{n} }{\sqrt{ \dot{n}^{2}}}-\xi .\dot{x} +\frac{1}{2} \xi^{2}.
\end{equation}

\subsubsection{The Symplectic Formalism of Modified Anyonic Model}
In the previous part, we added some gauge symmetry to the anyonic model artificially via \eqref{Lxi}. In this part we start the symplectic procedure to increase the gauge symmetries of the modified anyonic model \cite{kim}.

Corresponding momenta, which are calculated before as the relations \eqref{momenta0}, are now changed to the following relations, having three variables instead of two. 
\begin{align}
&   p_{\mu}=\frac{m \dot{n}_{\mu}}{\sqrt{ \dot{n}^{2}}}- \xi_{\mu}, \nonumber \\
&  \wp_{\mu}=m\frac{\dot{x}_{\mu}}{\sqrt{ \dot{n}^{2}}}-\frac{m\dot{n}^{2} \dot{x}_{\mu} }{( \dot{n}^{2})^{\frac{3}{2}}}, \nonumber \\
&   \pi_{\mu}=\dot{\xi}_{\mu} .
\end{align}
Calculating the canonical Hamiltonian, we have,
\begin{align}\label{Hc(0)}
& H_{c}= \dot{x}.p +\dot{n}.\wp +\dot{\xi}.\pi -L^{(0)}  \nonumber \\
& \quad = \frac{1}{2}  \pi^{2}\nonumber \\
& \quad =V^{(0)}.
\end{align}
Apparently, this system has the following primary constraints,
\begin{eqnarray}
&& \phi_{0}=n^{2} +1,\nonumber \\
&& \phi_{1}=(p+\xi )^{2}-m^{2}, \label{PSC1}\\
&& \phi_{2}=n.(p +\xi ),  \label{PSC2}\\
&& \phi_{3}=\wp.(p +\xi ) \label{PSC3}.
\end{eqnarray}
We observe that except the constraint \eqref{FC1}, all constraints in \eqref{PSC01} to  \eqref{PSC03} are modified. This shows that  \eqref{FC1} will be remained intact during the embedding formalism and thus, it can be ignored to the end of the process.

Now, by introducing  constraints $ \phi_{i} $ into the canonical sector of the first-order Lagrangian $L^{(0)}$, by means of the time derivative of Lagrange multipliers $\dot \lambda_{i} $, we get the first-iterative Lagrangian $L^{(1)}$  as,
\begin{align} L^{(1)}=\dot{x}.p+\dot{n}.\wp +\dot{\xi}.\pi -\sum_{i=1}^{3}\dot{\lambda}_{i} \phi^{i}-V^{(0)},
\end{align}
where, $ V^{(0)} $ is obtained in \eqref{Hc(0)}.

Applying symplectic formalism, one can obtain the existed secondary constraint of the model as,
\begin{align}
\phi_{4}=4\pi.(p+\xi)
\end{align}

It has been shown that one can construct the gauged Lagrangian by enlarging the phase space and adding a Wess-Zumino (WZ) term to the first order Lagrangian as,
\begin{equation}
\tilde{L}^{(1)}= L^{(1)}+ L_{WZ},
\end{equation}
where, $ L_{WZ} $ is a function depending on the original coordinates and WZ variable, defining with the help of two generators $ G^{(1)} $ and $ G^{(2)} $ \cite{NC4,8}.
\begin{equation}
L_{WZ}=G^{(1)}+G^{(2)},
\end{equation}
Checking the Poisson brackets of the constraints \eqref{PSC1} to \eqref{PSC3}, we see that $ \phi_{1} $ is first class with respect to the others. So, the generators $ G^{(1)} $ and $ G^{(2)} $ are defined by following relations \cite{NC4},
\begin{align}\label{Condition}
& G^{(1)}=\theta \phi_{4} \quad,  \nonumber \\
& G^{(2)}=-\theta^{2}\{\phi_{4},\phi_{1} \} .
\end{align}
In the above equations, $ \theta $ is the WZ variable, and its conjugate momentum, $ p_\theta $ which will not be appeared in the gauged model, is a first class constraint. Thus, it is the sign of the presence of the gauge symmetry in the obtained model.

Bringing this result into the first order Lagrangian, we obtain the gauged Lagrangian as,
\begin{align}\label{Lbar1}
 &\tilde{L}^{(1)}=\dot{x}.P +\dot{n}. \wp +\dot{\xi}.\pi +\sum_{i=1}^{3}\dot{\lambda}_{i}\phi^{i}-\frac{1}{2}\pi^{2}+4\pi (p+\xi )\theta -8(p+\xi )^{2}\theta^{2}.
\end{align}
Also, we can read off the embedded canonical Hamiltonian as,
\begin{equation}
 \tilde H_{c}=H_{c}-G^{(1)}-G^{(2)}.
\end{equation}

\section{Poisson structure of the gauged model}

Now, we calculate all constraints' corresponding momenta, using $ p_{\lambda i}=\frac{\partial \tilde{L}^{(0)}}{\partial \dot{\lambda}^{i}} $, to obtain $ \Phi_{i} = p_{\lambda i} $  (i=1,2,3), and $  \Phi_{4} =  p_{\theta} $.

Now, one can check out the consistency conditions,
\begin{equation}
0=\dot{\Phi}_{i}  =\{ \Phi_{i},\tilde{H}_{T} \},
\end{equation}
where,
\begin{align}
\tilde{H}_{T}=\tilde{H}_{c}+\lambda_{i}\Phi^{i}.
\end{align}

Due to the fact that  $ \{ \Phi_{i},\Phi_{j} \}=0$, secondary constraints will be obtained, calculating $ \Psi_{i}=\{\Phi_{i},\tilde{H}_{c} \} $. Thus,
\begin{align}
& \Psi_{1}=(P+\xi)^{2}-m^{2} \\
& \Psi_{2}=n.(P+\xi) \\
& \Psi_{3}=\wp.(P+\xi) \\
& \Psi_{4}=\pi .(p+\xi )+4m^{2}\theta
\end{align}

Now, calculating the consistency condition for $ \Psi_{i} $s, we obtain the other parts of constraints chain structure.
\begin{align}
& \Lambda_{1}=m^{2} \theta, \\
& \Lambda_{2}= n.\pi -\lambda_{3}m^{2}, \\
& \Lambda_{3}=\wp .\pi +\lambda_{2}m^{2}.
\end{align}
At this level the consistency will be terminated.
It is about time, we should consider one of the primary constraints which is left over during the symplectic procedure \eqref{FC1}. From its consistency condition, we have,
\begin{equation}
\Psi_{0}=\lambda_{3} \Lambda_{2}
\end{equation}

Now, we obtained all existed constraints of the model in the following chain structures.
\begin{align}\label{conschain}
& \phi_{0} \rightarrow \Psi_{0} \rightarrow\lambda_{3} \Lambda_{2} \nonumber \\ 
& \Phi_{1} \rightarrow \Psi_{1} \rightarrow \Lambda_{1} \nonumber \\
& \Phi_{2} \rightarrow \Psi_{2} \rightarrow \Lambda_{2} \nonumber \\ 
& \Phi_{3} \rightarrow \Psi_{3} \rightarrow \Lambda_{3} \nonumber \\ 
& \Phi_{4} \rightarrow \Psi_{4} 
\end{align}
It seems that we are in trouble, since we have encountered with the bifurcation in the first and third lines of \eqref{conschain}, which is not desirable. Thus, we should determine the branches that should be eliminated, in order to solve the problem of bifurcation. To overcome such a problem, we can propose two solutions.

First, If we try to eliminate $ \lambda_{3} $, i.e. we get $ \lambda_{3}=0 $, our chains will be as follows.
\begin{align}
& \phi_{0} \rightarrow \Psi_{0} \rightarrow \lambda_{3} \quad \qquad 
\textbf{which determines a coefficient} \nonumber \\ 
& \Phi_{1} \rightarrow \Psi_{1} \rightarrow \Lambda_{1}  \nonumber \\
& \Phi_{2} \rightarrow \Psi_{2} \rightarrow \Lambda_{2} \qquad \quad 
\textbf{which is in contradiction with} \quad \lambda_{3}=0\nonumber \\ 
& \Phi_{3} \rightarrow \Psi_{3} \rightarrow \Lambda_{3} \nonumber \\ 
& \Phi_{4} \rightarrow \Psi_{4} 
\end{align}
It is apparent that this contradicts our assumption.

Second and as a most proper approach, if we take $ \Psi_{0}=\lambda_{3}=0 $, then the canonical couple, $  \Phi_{3}=p_{\lambda3}$ and $\lambda_{3} $ , are constraints themselves and one can throw them away without considering them, calculating all Dirac brackets. We are allowed to do this elimination due to the fact that constraints which determine coefficients are second class constraints, while others are first class ones \cite{10}. Therefore, the chain structure, $\Phi_{3} \rightarrow \Psi_{3} \rightarrow \Lambda_{3} $, will have never been taken into the account, forever. Thus, in \eqref{conschain} we will have first, second, third and fifth lines, only.

According to this consideration, we would have more first-class constraints, because the Poisson brackets depending on the mentioned chain structure would remove easily.

Rewriting the changed constraint according to our consideration, we have,
\begin{equation}
\Lambda_{2}= n.\pi
\end{equation}
After extracting all possible constraints, using the simplified forms of them, we obtain only three first class constraints as, $ \phi_{0} ,\Phi_{1} $  and $\Phi_{2} $, which their Piosson brackets with others vanish. Also, we find six second class constraints in the model as, $ \Phi_{4},\Psi_{1},\Psi_{2},\Psi_{4},\Lambda_{1},\Lambda_{2} $. This classification can be done with the help of the algorithm introduced in \cite{10}. So, the corresponding Poisson brackets matrix of these second class constraints will be,
 \begin{equation} \label{deltaij}
\Delta =\left(
\begin{array}{cccccc}
 0 & 0 & 0 & -16 m^2 & -m^2 & 0 \\
 0 & 0 & 0 & 2 m^2 & 0 & 0 \\
 0 & 0 & 0 & 0 & 0 & 1 \\
 16 m^2 & -2 m^2 & 0 & 0 & 0 & 0 \\
 m^2 & 0 & 0 & 0 & 0 & 0 \\
 0 & 0 & -1 & 0 & 0 & 0
\end{array}
\right) 
 \end{equation}

 we will obtain the following Dirac brackets, which indicates the Poisson structure of the modified anyonic particle.
\begin{eqnarray*}
&& \{x_{\mu},x_{\nu}\}^{*} =\frac{1}{m^{2}}[(p_{\mu}+\xi_{\mu})\pi_{\nu}-(p_{\nu}+\xi_{\nu})\pi_{\mu}]  \\
&&  \{x_{\mu},\xi_{\nu}\}^{*} =-n_{\mu}n_{\nu}-\frac{1}{m^{2}}[(p_{\mu}+\xi_{\mu})(p_{\nu}+\xi_{\nu})] \\
&& \{x_{\mu},p_{\nu}\}^{*} = \delta_{\mu\nu}  \\  
&&  \{\wp_{\mu},\wp_{\nu}\}^{*} =(p_{\mu}+\xi_{\mu})\pi_{\nu}-(p_{\nu}+\xi_{\nu})\pi_{\mu}  \\
&&  \{x_{\mu},\wp_{\nu}\}^{*} =n_{\mu}\pi_{\nu}  \\
&&  \{n_{\mu},\wp_{\nu}\}^{*} =\delta_{\mu\nu}  \\
&&  \{\xi_{\mu},\wp_{\nu}\}^{*} =n_{\mu}(P_{\nu}+\xi_{\nu}) \\
&&  \{\pi_{\mu},\pi_{\nu}\}^{*} =\frac{1}{m^{2}}[(p_{\mu}+\xi_{\mu})\wp_{\nu}-(p_{\nu}+\xi_{\nu})\wp_{\mu}] \\
&&   \{x_{\mu},\pi_{\mu}\}^{*} =\frac{-1}{m^{2}}[(p_{\mu}+\xi_{\mu})\pi_{\nu}+(p_{\nu}+\xi_{\nu})\pi_{\mu}] \\
&&  \{\xi_{\mu},\pi_{\nu}\}^{*} =\delta_{\mu\nu}+n_{\mu}n_{\nu}+\frac{1}{m^{2}}[(p_{\mu}+\xi_{\mu})(p_{\nu}+\xi_{\nu}) ]\\
&&   \{\wp_{\mu},\pi_{\nu}\}^{*} =n_{\mu}\pi_{\nu} \\
&&  \{x_{\mu},p_\theta\}^{*} =-12 (p_{\mu} + \xi_{\mu})  \\
&&  \{\pi_{\mu},p_\theta\}^{*} =12 (p_{\mu} + \xi_{\mu})  \\
&&   \{\theta,p_\theta\}^{*} =1 
\end{eqnarray*}
We see that in comparison with \eqref{Poisson0}, the modified anyonic particle is remained noncommutative, but with a more extended phase space \cite{Meljanac}.
\section{Physical Degrees of Freedom}

The number of physical degrees of freedom can be obtained with the help of following relation,
\begin{equation}
\mathcal{N}=\#( \mathcal{Q}_{i}-\mathcal{FC}_{i}-\frac{1}{2}\mathcal{SC}_{i}),
\end{equation}
where, $ \mathcal{Q}_{i} $ is the number of coordinates, and $ \mathcal{FC}_{i} $ and $\mathcal{SC}_{i} $ are the numbers of first class and  second class constraints respectively \cite{Henneaux2}. 

In a $ (2+1) $ dimensional spacetime, the number of physical degrees of freedom for original model is,
\begin{equation}\label{n1}
\mathcal{N}_{Original}=6-2-1=3,
\end{equation}
while, for the modified model it will be obtained as,
\begin{equation}\label{n2}
\mathcal{N}_{Modified}=12-3-3=6.
\end{equation}

Aparently, the gauged model has three extra physical degrees of freedom in comparison with the original model, which can be interpreted as the interaction of the particle with the electromagnetic (gauge) field \cite{chai1,chai2,chai3,chai4,jing}.

\section{Gauge Transformations' Generators}
The generators of infinitesimal gauge transformations can be obtained with the help of Poisson brackets of the first class constraints and the phase space coordinates of the Lagrangian \eqref{Lbar1}, i.e. $ \chi_{\alpha} $ , via the following relation \cite{Henneaux2,Shirzad},
\begin{equation}\label{Shirzad henneaux}
 \delta \chi^{(1)}_{ \alpha}=\{\chi^{(1)}_{ \alpha},\phi_{j}\}\epsilon^{j}.
\end{equation}
where,  $ \epsilon_{i} $ are infinitesimal time dependent parameters, and $ \phi_{i}  $ are first-class constraints .
So, the the infinitesimal gauge transformations of the modified anyonic particle model which determine its gauge symmetrise are as follows,
\begin{align}\label{Gauge transf}
\begin{array}{ll}
  \delta x_{\mu}=\textbf{0}, 	 & \qquad \delta p_{\mu}=\textbf{0}, \\
  \delta n_{\mu}=\textbf{0}, & \qquad \delta \wp_{\mu}=-2n_{\mu}\epsilon_{0}, \\
  \delta \xi_{\mu}= \textbf{0}, & \qquad \delta \pi_{\mu}=\textbf{0},  \\
  \delta \lambda_{1}=\epsilon_{1},\\
  \delta \lambda_{2}=\epsilon_{2},\\
  \delta \theta=0 .
\end{array}
\end{align}
Apparently, the Lagrangian \eqref{Lbar1} and the corresponding Hamiltonian are invariant under these transformations.

\section*{Conclusion}
In relativistic classical mechanics point of view, the basic Lagrangian of the anyons are constructed with the help of some nonphysical coordinates, such as $ n_{\mu} $. The corresponding angular momenta of these coordinates make any spin for the particle in quantization process. Classically, relating physical degrees of freedom of particle to its quantum spin is an unsolved problem.

 Our tool to investigate this problem is gauging via the symplectic procedure. We observe that the nonphysical constraint, i.e. $ n^{2}+1=0 $ will not be affected. So, the gauged model of anyons remains anyonic. But, in this manner, we propose another procedure, like BFT \cite{BFFT1, BFFT2, BFFT3,Monem2, BFFT4} with which we are working now, within which we predict that the mentioned constraint will be affect. Hence, after gauging the anyon's Hamiltonian, we derive an interacting model.

Dirac brackets of the new gauged model show that the new model is noncommutatitve after quantization is applied \cite{Huang}.

Moreover, as we mentioned before, we find some correction terms in our results, which can be considered as  interaction terms of anyon and electromagnetic (gauge) field. Extracting and comparing these terms with others \cite{chai1,chai2,chai4}, will be done with the help of constraints equations  in future.

At the end, by eliminating new variables, we witness that none of them is remained in the new model, and counting physical degrees of freedom of two models, confirms our claim (See \eqref{n1} and \eqref{n2}).

\end{document}